\documentstyle[sprocl]{article}

\def\al{\alpha}
\def\be{\beta}
\def\ga{\gamma}

\def\et{\eta}

\def\ka{\kappa}
\def\la{\lambda}

\def\cL{{\mathcal L}}

\def\half{{\textstyle{1\over 2}}}

\def\frac#1#2{{\textstyle{{#1}\over {#2}}}}

\def\lsim{\mathrel{\rlap{\lower4pt\hbox{\hskip1pt$\sim$}}
    \raise1pt\hbox{$<$}}}
\def\gsim{\mathrel{\rlap{\lower4pt\hbox{\hskip1pt$\sim$}}
    \raise1pt\hbox{$>$}}}
\def\sqr#1#2{{\vcenter{\vbox{\hrule height.#2pt
         \hbox{\vrule width.#2pt height#1pt \kern#1pt
         \vrule width.#2pt}
         \hrule height.#2pt}}}}

\def\lvb#1#2{e_{#1#2}}

\newcommand{\beq}{\begin{equation}}
\newcommand{\eeq}{\end{equation}}
\newcommand{\bea}{\begin{eqnarray}}
\newcommand{\eea}{\end{eqnarray}}
\newcommand{\bit}{\begin{itemize}}
\newcommand{\eit}{\end{itemize}}
\newcommand{\rf}[1]{(\ref{#1})}

\begin{document}

\title{Nambu-Goldstone and Massive Modes in
Gravitational Theories with Spontaneous Lorentz Breaking}

\author{R.\ Bluhm}

\address{Physics Department, \\
Colby College, \\ 
Waterville, ME 04901, USA\\ 
E-mail: rtbluhm@colby.edu}

\maketitle

\abstracts{
Spontaneous breaking of local Lorentz symmetry is of interest as a 
possible mechanism originating from physics at the Planck scale.
If such breaking occurs, however, it raises the questions of what
the fate of the Nambu-Goldstone modes is, whether a Higgs
mechanism can occur, and whether additional massive modes
(analogous to the Higgs particle) can appear.
A summary of some recent work looking at these questions
is presented here.
}

\section{Introduction}	

The idea that Lorentz symmetry might be spontaneously broken
began to catch on when it was shown that mechanisms
in string theory might lead to this form of symmetry breaking.\cite{ks}
Since then, spontaneous Lorentz breaking has been examined in
its own right in a number of contexts,
including investigating its phenomenological effects and
its effects on gravity.
However,
as soon as a theory allows spontaneous breaking of a symmetry,
well-known consequences from particle physics 
must be considered and addressed.
The first is the Goldstone theorem,
which states that when a continuous symmetry is
spontaneously broken, massless Nambu-Goldstone (NG) modes appear.
The second is the possibility of a Higgs mechanism,
resulting in massive gauge fields,
for the case when the symmetry is local.
The third is the possibility that additional massive modes might appear
(analogous to the Higgs boson in the case of the electroweak model).
Clearly,
all three of these can have physical implications and must be
accounted for in any theory with spontaneous symmetry breaking.

In this work,
these processes are examined for the case where it is
Lorentz symmetry that is spontaneously broken.\cite{ks,akgrav,rbak,rbffak} 
First, the fate of the NG modes is examined.
Then, since Lorentz symmetry is a local symmetry in
the context of gravity,
the possibility of a Higgs mechanism is considered.
Lastly,
the possibility of additional massive modes 
(analogous to the Higgs particle) is considered as well.
An explicit illustration of these processes is
given for the case of a bumblebee model,
in which a vector field acquires a nonzero vacuum value.

\section{Spontaneous Lorentz Breaking}

In a gravitational theory,
Lorentz symmetry acts in local frames, 
transforming tensor components with respect to a local basis, 
e.g., $T_{abc}$
(where Latin indices denote components with respect to a local frame).
Similarly,
diffeomorphisms act in the spacetime manifold, 
transforming components with respect to the spacetime coordinate system,
e.g., $T_{\lambda\mu\nu}$ (denoted using Greek indices).
These local and spacetime tensor components are linked by a vierbein.
For example,
the spacetime metric and local Minkowski metric are related by
\begin{equation}
g_{\mu\nu} = e_\mu^{\,\,\, a} e_\nu^{\,\,\, b} \eta_{ab}  .
\label{vier}
\end{equation}       

With a vierbein formalism, 
spinors can naturally be incorporated into a theory.  
A vierbein formalism also parallels gauge theory,
with Lorentz symmetry acting as a local symmetry group.
The spin connection $\omega_\mu^{\,\,\, ab}$ enters in covariant
derivatives that act on local tensor components and plays the role of
the gauge field for the Lorentz symmetry.
In contrast,
the metric excitations,
e.g., $h_{\mu\nu} = g_{\mu\nu}  - \et_{\mu\nu}$,
act as the gauge
fields for the diffeomorphism symmetry.
In the context of a vierbein formalism,
there are primarily two geometries that can be distinguished.
In a Riemannian geometry (with no torsion), 
the spin connection is nondynamical and does not propagate.  
However, in a Riemann-Cartan geometry (with nonzero torsion), 
the spin connection must be treated as independent degrees of freedom
that in principle can propagate.

Local Lorentz symmetry is spontaneously broken when a local tensor 
field acquires a nonzero vacuum expectation value (vev),
e.g.,
for the case of a three-component tensor,
\begin{equation}
<T_{abc}> \, = t_{abc} .
\label{Tvev}
\end{equation}       
Spontaneous Lorentz breaking can be introduced into a theory
dynamically by adding a potential term $V$ to the Lagrangian.
For example, a potential of the form
\beq
V \sim(T_{\la\mu\nu} \,
g^{\la\al} g^{\mu\be} g^{\nu\ga}  \,
T_{\al\be\ga}
\pm \, t^2)^2 ,
\label{VT2}
\eeq
consisting of a quadratic function of products of the tensor
components $T_{\la\mu\nu}$,
has a minimum when \beq
T_{\la\mu\nu} \, g^{\la\al} g^{\mu\be} g^{\nu\ga} \,
T_{\al\be\ga} = \mp \, t^2 .
\label{condT}
\eeq
Solutions of Eq.\ \rf{condT} span a degenerate space of
possible vacuum solutions.
Spontaneous Lorentz breaking occurs when a
particular vacuum value $t_{abc}$ in the local frame is chosen,
obeying $\mp t^2 = t_{abc} \, \et^{pa} \et^{qb} \et^{rc} \, t_{pqr}$,
where the sign depends on the timelike or spacelike nature of the tensor.

\section{Nambu-Goldstone Modes}

Consider a theory with a tensor vev in a local Lorentz frame,
$<T_{abc}> \, = t_{abc}$,       
which spontaneously breaks Lorentz symmetry.  
Since the vacuum value for the vierbein is also a constant or fixed function, 
e.g., $<e_\mu^{\,\,\, a}> \, = \delta_\mu^{\,\,\, a}$,       
the spacetime tensor therefore has a vev as well,
\begin{equation}
<T_{\lambda\mu\nu}> \, = t_{\lambda\mu\nu} .
\label{Tmunuvev}
\end{equation}     
This means that when Lorentz symmetry is spontaneously broken,
diffeomorphisms are spontaneously broken as well.
This implies that NG modes should appear (in the absence of a Higgs mechanism)
for both of these broken symmetries.
In general,
the NG modes consist of field excitations that stay within
the minimum of the potential $V$.
They therefore obey the condition \rf{condT}.
A solution of this condition is given in terms of the vierbein
and the local vev, 
\beq
T_{\la\mu\nu} = \lvb \la a \lvb \mu b \lvb \nu c \, t^{abc} .
\eeq

As a general rule,
there can be up to as many NG modes as there are broken symmetries. 
Since the maximal case corresponds to six broken Lorentz generators 
and four broken diffeomorphisms,
there can therefore be up to ten NG modes.
Where do the NG modes reside?
In general, the answer depends on the choices of gauge.  
However, one natural choice is to put all of the NG modes into the vierbein.
A counting argument shows this is possible.
The vierbein $e_\mu^{\,\,\, a}$ has 16 components.  
With no spontaneous Lorentz violation, 
typically the six Lorentz and four diffeomorphism degrees
are used to gauge away ten components,
leaving up to six independent degrees of freedom.  
(Note that a general gravitational theory can have up to
six propagating metric modes,
but general relativity is special in that there are only two). 
In contrast, in a theory with spontaneous
Lorentz breaking, 
up to all ten NG modes can potentially propagate
as additional degrees of freedom in the vierbein.

\section{Gravitational Higgs Mechanisms}

With two sets of broken symmetries,
local Lorentz transformations and diffeomorphisms,
there are potentially two types of  Higgs mechanisms. 
Furthermore,
there is the possibility that additional massive modes can
exist as excitations that do not stay in the minimum of
the potential $V$.

For the case of the broken diffeomorphisms.
it was shown that the conventional Higgs mechanism 
involving the metric does not occur.\cite{ks}  
This is because the mass term that is generated by covariant derivatives
involves the connection,  which consists of derivatives of the metric
and not the metric itself.
As a result,
no mass term for the metric is generated 
according to the usual Higgs mechanism.
However, it was also shown that because of the form of the potential, 
e.g., as in Eq.\ \rf{VT2}, 
quadratic terms involving the metric can arise.
This results in an alternative form of the Higgs mechanism\cite{ks}
that has no direct analogue in nonabelian gauge theory.
(In nonabelian gauge theory, 
the potential $V$ involves only the scalar Higgs fields
and not the gauge fields.
In contrast here,
both the metric and tensor fields enter in the massive-field excitations).
The additional mass terms that arise in this alternative Higgs mechanism
can potentially modify gravity in a way that avoids the
van Dam, Veltmann, and Zakharov discontinuity.\cite{vdvz}
They are therefore potentially interesting in studies of
modified gravity theory.

In contrast, for the case of the broken Lorentz symmetry, 
it is found that a conventional Higgs mechanism can occur.\cite{rbak}
In this case, 
the relevant gauge field is the spin connection.  
This field appears directly in covariant derivatives acting on local tensor
components,
and for the case where the local tensors acquire a vev,
quadratic mass terms for the spin connection can be generated.
However, a viable Higgs mechanism involving the spin connection can 
occur only if the spin connection is a dynamical (i.e., propagating) field.  
This then requires that there is nonzero torsion and 
that the geometry is Riemann-Cartan.
As a result,
a conventional Higgs mechanism for the spin connection is possible, 
but only in a Riemann-Cartan geometry.
However,
even if torsion is permitted,
constructing a viable model with a massive propagating spin connection
that is ghost- and tachyon-free remains a challenging and open problem.\cite{rbak}
Therefore,
for simplicity in the remainder of this work,
a Riemann spacetime (with no torsion) is assumed.
In this restricted context,
the only possible process giving rise
to massive modes is the alternative Higgs mechanism,
in which massive modes are due to excitations that
do not stay in the minimum of the potential $V$.

\section{Bumblebee Models}

To investigate further the effects of NG and massive
modes in theories with spontaneous Lorentz violation,
it is useful to work in the context of a definite model. 
The simplest example involves a vector field 
with a nonzero vev.
Models of this type are known as bumblebee models.\cite{ks,akgrav}
Examples have been studied in various forms by a number of 
authors.\cite{ks,akgrav,rbak,rbffak,kl01,baak05,kb06,ejm,kt02,mof03,grip04,cl04,bp05,ems05,acfn,clmt06}

Bumblebee models are defined as field theories with
a vector field $B_\mu$ that acquires a nonzero vev,
$<B_\mu> \, = b_\mu$.    
The vev is induced by a potential $V$ in the Lagrangian that has a minimum 
when the vacuum solution holds.
Bumblebee models can be defined with generalized kinetic terms
for the vector and gravitational fields.
However,
for brevity, 
an example with a Maxwell kinetic term 
is considered here.
The Lagrangian then has the form
${\cL} = {\cL}_{\rm G} + {\cL}_{\rm B} + {\cL}_{\rm M}$,
where ${\cL}_{\rm G}$
describes the pure-gravity sector,
${\cL}_{\rm M}$ describes the matter sector
(including possible interactions with $B_\mu$),
and 
\begin{equation}
{\cL}_{\rm B} =  - \frac 1 4 B_{\mu\nu} B^{\mu\nu} - V(B_\mu B^\mu \pm b^2)  ,
\label{BBL}
\end{equation}
describes the bumblebee field.
(For simplicity, additional possible interactions between the
curvature tensor and $B_\mu$ are neglected here as well).
The bumblebee field strength in Riemann spacetime is
$B_{\mu\nu} = \partial_\mu B_\nu - \partial_\nu B_\mu$.

A noteworthy feature of all bumblebee models is that
they do not have local $U(1)$ gauge symmetry.
This symmetry is broken explicitly by the presence of the potential $V$.
However,
it is common to include couplings to matter that involve
the notion of charge in the matter sector.
For example,
terms involving current couplings with charged matter
can be included by defining,
$\cL_{\rm M} = B_\mu J^\mu$ with $D_\mu J^\mu = 0$.
In this case, the theory has a global $U(1)$ symmetry that
gives rise to charge conservation in the matter sector.
This assumption also implies that initial values can be chosen
that maintain stability of the Hamiltonian.\cite{rbffak}

Different forms of the potential $V$ can be considered.
One example is a smooth quadratic potential,
\beq
V = \half \ka (B_\mu B^\mu \pm b^2)^2 ,
\label{Vkappa}
\eeq
where $\ka$ is a constant (of mass dimension zero).
This type of potential allows both NG excitations 
(that stay within the potential minimum)
as well as massive excitations
(that do not).
An alternative would be to consider
a linear Lagrange-multiplier potential
\beq
V = \la (B_\mu B^\mu \pm b^2) , 
\label{Vsigma}
\eeq
where the Lagrange-multiplier field $\la$ imposes a constraint
that only allows NG excitations in $B_\mu$ and
excludes massive-mode excitations.
However,
for definiteness here,
the smooth potential \rf{Vkappa} is chosen,
which allows a massive-mode excitation.

For such a bumblebee model,
three Lorentz symmetries and one diffeomorphism are broken.
Therefore, up to four NG modes can appear.  
However, the diffeomorphism NG mode 
is found not to propagate.\cite{rbak}
It drops out of the kinetic terms and is purely an auxiliary field.  
In contrast, the Lorentz NG modes do appear in the form of
two massless transverse modes and one auxiliary mode.
These have properties similar to the photon in electrodynamics,
which raises the interesting possibility that photons might be described
as NG modes in theories with spontaneous Lorentz violation.\cite{rbak}
Previous links between QED gauge fields, fermion composites, and NG modes 
have been uncovered in flat spacetime (with global Lorentz symmetry).\cite{ngphoton}  
However,
bumblebee models are different.
They consist of theories with a noncomposite vector field, 
have no local U(1) gauge symmetry,
and give rise to photons as NG modes in the presence of gravity.
Note that bumblebee models also include possible couplings between
the vacuum value $b_\mu$ and a matter current $J^\mu$.
Such an interaction can provide an unmistakable signature of
physical Lorentz violation that would distinguish it
from any gauge-fixed form of QED.
Note as well that any such signal would be contained
in the Standard-Model Extension(SME).\cite{sme}
Thus,
on-going investigations of Lorentz breaking using the SME
have sensitivity to all signals of spontaneous Lorentz breaking
involving couplings between matter and the background vevs.

To determine more thoroughly whether conventional Einstein-Maxwell
solutions can emerge from bumblebee models,
the role of the massive mode must be investigated.\cite{rbffak}
This mode constitutes an additional degree of freedom beyond those
of the NG modes.
It also alters the form of the initial-value problem.
For simplicity,
only the case of a purely timelike vacuum vector
$b_\mu = (b,0,0,0)$ is considered here.
In this case,
in the weak-field limit,
it is found that the massive mode does not propagate as a free field.
Instead,
it remains purely an auxiliary field that has no time dependence.
As a result,
its value is fixed by the initial conditions at $t=0$.
Although it does not propagate,
the massive mode can nevertheless alter the form of the static potentials.
An example of this can be seen by solving for the
modified static potentials in the presence of a point particle
with mass $m$ and charge $q$.
It is found that both the electromagnetic and gravitational
potentials are modified by the presence of the massive mode,
where the specific forms of the modified potentials depend on the
assumed initial value of the static massive mode.
There are therefore numerous cases that could be explored,
including examples that might be relevant in considering
alternative explanations of dark matter.
However,
in the large-mass limit
(e.g., approaching the Planck scale),
excitation of the massive mode is highly suppressed,
and the static potentials approach the conventional Coulomb 
and Newtonian forms.
In the limit of a vanishing massive mode,
these become exact expressions.
As a result,
it is found that the usual Einstein-Maxwell solutions
(describing both propagating photons and the usual static potentials)
can emerge from a bumblebee model (without local U(1) symmetry),
in which local Lorentz symmetry is spontaneously broken.

\section*{Acknowledgments}

This work was supported  
by NSF grant PHY-0554663.

\end{document}